\documentclass[aps,prd,showpacs,11pt,tightenlines,preprint,preprintnumbers,superscriptaddress,nofootinbib]{revtex4-1}

\usepackage[colorlinks=true,
urlcolor=cambridgeblue,
linkcolor=darkraspberry,
citecolor=cambridgeblue,
linktocpage=true,
pdfproducer=medialab,
pdfa=true,
anchorcolor=blue]{hyperref}% 
\usepackage{xcolor}
\definecolor{cambridgeblue}{rgb}{0.64, 0.76, 0.68}
\definecolor{darkraspberry}{rgb}{0.53, 0.15, 0.34}
\usepackage{graphicx}% Include figure files
\usepackage{dcolumn}% Align table columns on decimal point
\usepackage{bm}% bold math
\usepackage{hyperref}% add hypertext capabilities
\usepackage{amssymb}
\usepackage{placeins}
\usepackage[utf8]{inputenc}
\usepackage{float}
\usepackage[utf8]{inputenc} 
\usepackage{xcolor}
\usepackage{feynmp-auto}

\begin{document}

\preprint{IFIC/19-06}

\title{Revisiting the LHC reach in the displaced region of the minimal left-right symmetric model }

\author{Giovanna Cottin}
\email{gcottin@phys.ntu.edu.tw}
\affiliation{Department of Physics, National Taiwan University, Taipei  10617, Taiwan}

% I prefer to list only one corresponding author to avoid lot's of fwd to me 

\author{Juan Carlos Helo}
% \email{jchelo@userena.cl }
  \affiliation{
Departamento de F\' isica, Facultad de Ciencias, Universidad de La Serena, 
Avenida Cisternas 1200, La Serena, Chile}

\author{Martin Hirsch}
% \email{mahirsch@ific.uv.es}
\affiliation{AHEP Group, Instituto de F\'{\i}sica Corpuscular --
    CSIC/Universitat de Val{\`e}ncia, 
    Edificio de Institutos de Paterna, Apartado 22085,
    E--46071 Val{\`e}ncia, Spain}

\author{Dario Silva}
% \email{darioss90@gmail.com}
  \affiliation{
Departamento de F\' isica, Facultad de Ciencias, Universidad de La Serena, 
Avenida Cisternas 1200, La Serena, Chile}

\date{\today}

%%%%%%%%%%%%%%%%%%%%%%%%%%%%%%%%%%%%%%%%%%%%%%%%%%%%%%%%%%%%%%%%%%%%

\begin{abstract}
We revisit discovery prospects for a long-lived sterile neutrino $N$
at the Large Hadron Collider (LHC) in the context of left-right
symmetric theories. We focus on a displaced vertex search strategy
sensitive to $\mathcal{O}$(GeV) neutrino masses produced via a
right-handed $W_{R}$ boson. Both on-shell and off-shell Drell-Yan
production of $W_{R}$ are considered. We estimate the reach as a
function of $m_{N}$ and $m_{W_{R}}$. With $\sqrt{s}=13$ TeV and 300/fb
of integrated luminosity, the LHC can probe neutrino masses as high as
$\sim 30$ GeV and $m_{W_{R}}$ around 6 TeV. The reach goes up to
11.5 TeV with 3000/fb and $m_{N}\sim 45$ GeV. This represents an
improvement of a factor of 2 in sensitivity with respect to earlier
work.
\end{abstract}

\maketitle

\section{Introduction}

The origin of the mechanism responsible for providing mass to the neutrinos in the Standard Model (SM) remains unknown. Left-right symmetric theories~\cite{Pati:1974yy,Mohapatra:1974gc} are popular candidates to explain light neutrino masses via the so-called see-saw mechanism~\cite{Minkowski:1977sc}, that additionally address the origin of parity violation in the SM weak sector.  

In the minimal left-right framework, parity is broken spontaneously together with the new $SU(2)_{R}$ right-handed weak interaction, generating Majorana masses for both light and right-handed neutrinos $N$. Production and decay of the right-handed or sterile neutrino $N$ depends mostly on the unknown mass of the new, heavy right-handed gauge boson, $W_{R}$. For $m_{N}\ll m_{W_{R}}$, the golden channel is the Keung-Sejanovi\'c process (KS)~\cite{Keung:1983uu} where production proceeds through a $W_{R}$ via $pp\to W^{\pm}_{R} \to l^{\pm} N \to l^{\pm}l^{\pm}jj$. 

Recently the authors in Ref.~\cite{Nemevsek:2018bbt} have
systematically classified all the signatures resulting from the KS
process into four regions, based on the mass of the sterile neutrino
$N$. These regions are: {\it{standard}}, in which $m_{N}\gtrsim200$
GeV, leading to the well studied same sign leptons plus jets ($lljj$)
signature~\cite{Keung:1983uu} (for recent reviews, 
see Refs.~\cite{Deppisch:2015qwa,Cai:2017mow} and references therein); the {\it{merged}}
region in which the final state is a prompt lepton and a neutrino jet
($lj_{N}$); the {\it{displaced}} region where the neutrino appears at
a visible distance from the primary vertex leading to a displaced
vertex signature; and the {\it{invisible}} region where $N$ escapes
the detector and leads to events with missing transverse momenta.

Several experimental searches at the LHC target the KS process in the {\it{standard}} region, in a regime where $N$ decays promptly. Most recently, the ATLAS~\cite{Aaboud:2018spl} and CMS~\cite{Sirunyan:2018pom} collaborations provide stringent limits on the mass of $W_{R}$, for $N$ masses close to the TeV scale.  $W_{R}$ masses below $\sim 3.5$ TeV are already excluded by dijet resonance searches~\cite{Aaboud:2017yvp}. Existing searches for a new heavy boson decaying to leptons and missing transverse momenta~\cite{Khachatryan:2016jww,Aaboud:2017efa} can constrain part of the {\it{invisible}} region~\cite{Nemevsek:2018bbt}. No dedicated experimental searches targeting the {\it{merged}} and {\it{displaced}} regions exists yet, when $N$ has a mass below the electroweak scale~\footnote{Although we note that the CMS search for right-handed neutrinos in events with three prompt charged leptons in the final state~\cite{Sirunyan:2018mtv} is sensitive to $N$ masses below $40$ GeV, where the focus is on a simplified model that extends the SM fermionic sector with only one right-handed neutrino.}.

For sterile neutrino masses below the electroweak scale, $N$ becomes a long-lived particle that can be detected via its displaced decay to a lepton and jets~\cite{Helo:2013esa,Nemevsek:2011hz}. There is a rapidly growing interest in the study of the LHC capabilities to those heavy neutrino signatures, and heavy neutral leptons in general~\cite{Abada:2018sfh} (see also~\cite{Alimena:2019zri} for a community white paper). Long-lived particle signatures in left-right symmetric theories include displaced diphoton jets~\cite{Dev:2016vle}, displaced jets~\cite{Nemevsek:2018bbt}, displaced neutrino jets~\cite{Mitra:2016kov} and displaced vertices~\cite{Helo:2013esa,Castillo-Felisola:2015bha,Castillo-Felisola:2015bha,Dev:2017dui,Cottin:2018kmq}. 

The purpose of this paper is to revisit the sensitivity estimates we made in~\cite{Cottin:2018kmq} for the {\it{displaced}} KS region. In~\cite{Cottin:2018kmq}, we focused on the on-shell production of $W_{R}$. The authors in~\cite{Nemevsek:2018bbt} have demonstrated that for $m_{W_{R}}> 5$ TeV, the KS process is dominated by off-shell production via $W^{*}_{R}$ (see Figure 3 of~\cite{Nemevsek:2018bbt}). Considering this contribution will give an enhancement in the cross-section and therefore in sensitivity of our displaced strategy. Note that in~\cite{Cottin:2018kmq} we considered a zero background, multitrack displaced vertex search strategy, as opposed to the recognition of displaced jets performed in~\cite{Nemevsek:2018bbt}, which is sensitive to higher neutrino masses. We compare and discuss on the complementarity of our strategy and overlaps with the other KS regions in the regime $m_{N}<100$ GeV.

\section{Simulations of $W_{R}$ production and $N$ decay}
\label{LRtheory}

The minimal left-right symmetric extension of the SM~\cite{Pati:1974yy,Mohapatra:1974gc,Mohapatra:1980yp} has gauge group $SU(2)_{L}\times SU(2)_{R}\times U(1)_{B-L}$. This model contains a right-handed gauge boson $W_{R}$ and three right-handed Majorana neutrinos, with lightest state $N$. In what follows, we consider only one neutrino in the kinematic region of interest ($m_{N}\ll m_{W_{R}}$, and $m_{N}<100$ GeV) and we restrict, for simplicity, our discussion to sterile neutrino mixing with the electron sector only\footnote{We do not expect our results to change significantly if flavour mixing in the muon sector is present, as reconstruction efficiencies are rather similar. The sensitivity can be smaller when tau mixing is present due to the difficulties in reconstructing tau leptons~\cite{Cottin:2018nms}.}.

Drell-Yan production of $W_{R}$ proceeds via $pp\to W^{\pm}_{R} \to l^{\pm} N$, with $N$ further decaying displaced to $N\to l^{\pm} jj$. This production may be dominantly off-shell~\cite{Ruiz:2017nip,Nemevsek:2018bbt}. The authors in~\cite{Nemevsek:2018bbt} have shown that the off-shell enhancement can be significant for light $m_{N}$. They also provide a dedicated event generator, the KS Event Generator ({\textsc{KSEG}})~\cite{KSEG}, that considers off-shell $W_{R}$ production as well as light (or heavy) $N$. It also appropriately deals with narrow resonances, as is the case of a long-lived $N$. This is explained in Appendix C of~\cite{Nemevsek:2018bbt}.

We generate parton-level events with {\textsc{KSEG}}~\cite{KSEG} for a grid of masses covering $m_{N}=(10,50)$ GeV and $m_{W_{R}}=(2,12)$ TeV. The $N$
lifetimes are computed numerically as described in Appendix A of~\cite{Nemevsek:2018bbt}. The KS event generator takes special care in simulating the case of a long-lived neutrino, and LHE events are generated with decay vertex information. The NLO corrections to the cross-section are taken into account by considering a constant $k-$factor of 1.3, as calculated in~\cite{Mitra:2016kov}. We further hadronize events from {\textsc{KSEG}} with {\textsc{Pythia 8}}~\cite{Sjostrand:2014zea} from where we can retain all the relevant information, (including lifetime) for our analysis. 

The $N$ charged decay products (i.e tracks from hadronization) are identified to come from a common displaced vertex. A detailed detector response as a function of the displaced vertex invariant mass and number of tracks is applied to each vertex passing the selection criteria~\cite{Cottin:2018kmq}, described in the following section.

\section{Displaced vertex reach}
\label{sec:SimResults}

The displaced vertex search strategy used in this article is the same one proposed in our earlier works in Refs.~\cite{Cottin:2018kmq,Cottin:2018nms}\footnote{With the exception of the ``trackless jet cut" defined in~\cite{Cottin:2018kmq}. We originally kept this cut to follow as closely as possible the selections made in the ATLAS search~\cite{Aad:2015rba,Aaboud:2017iio}. This cut is not necessary and has no effect in the limits presented in this work.}, which is inspired by the ATLAS zero background search in~\cite{Aaboud:2017iio}. Our search is sensitive to decays occurring in the inner trackers of the LHC detectors. Events are triggered by a prompt electron (reconstructed as in~\cite{Cottin:2018nms}) with $p_{T}>25$ GeV, and additional cuts are imposed in the selection of the displaced tracks and vertex. 

The displaced vertices are selected by reconstructing tracks with transverse impact parameter $|d_{0}|$ bigger than 2 mm, and with $p_{T} > 1$ GeV. The vertex position must be within 4 mm and 300 mm (tracker acceptance), and must have more than three tracks. The invariant mass of the displaced vertex (which is calculated assuming all tracks have the mass of the pion) is required to be $\geq5$ GeV. As in Refs.~\cite{Cottin:2018kmq,Cottin:2018nms}, we quantify the detector response to displaced vertices by making use of the ATLAS parametrized efficiencies given in~\cite{Aaboud:2017iio}.

\begin{figure}[h]
\centering
\includegraphics{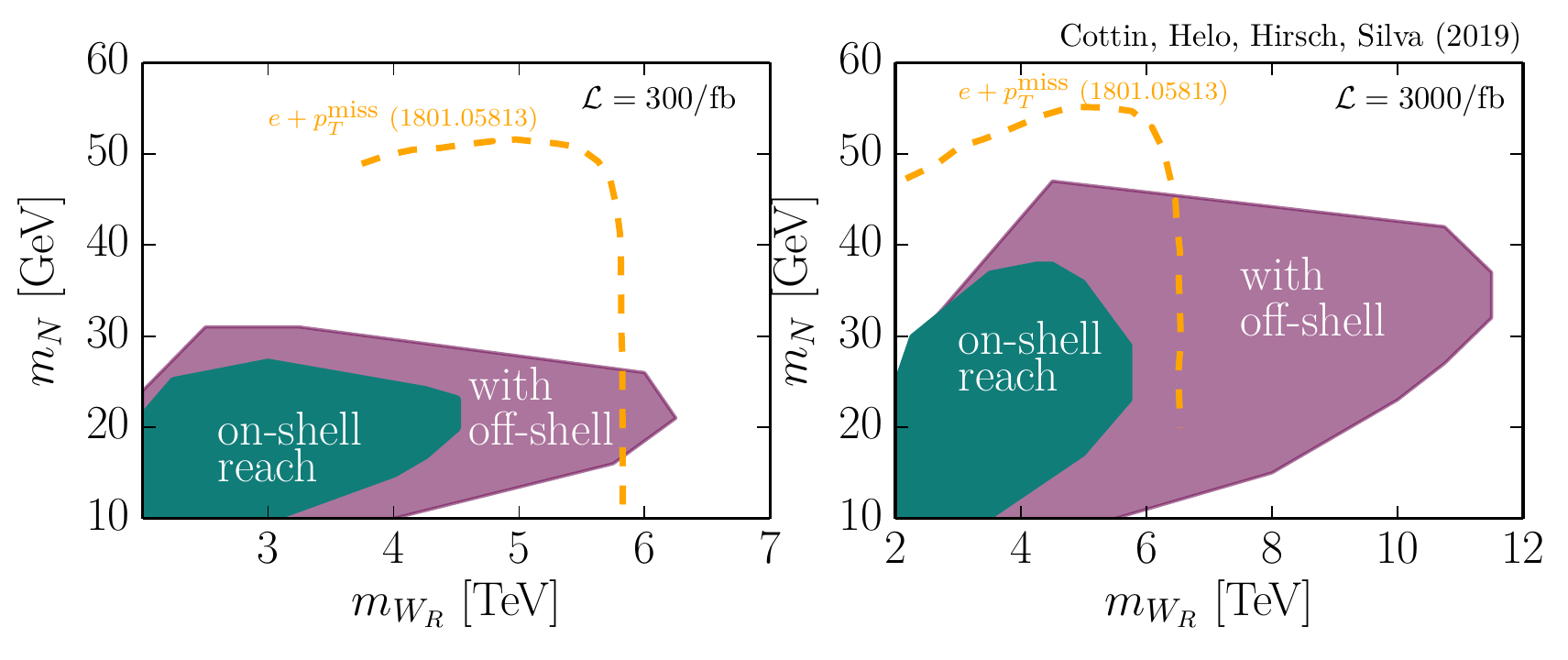}
\caption{$95\%$ CL reach in the $(m_{W_{R}},m_{N})$ plane at
  $\sqrt{s}=13$ TeV of our proposed displaced strategy for
  $\mathcal{L}=300$/fb (left) and $\mathcal{L}=3000$/fb
  (right). Comparison of the purely on-shell~\cite{Cottin:2018kmq} (green region) and considering off-shell $W_{R}$ production (purple) are shown. The projected exclusions in 
  the $e + p^{\mbox{miss}}_{T}$ channel from~\cite{Nemevsek:2018bbt} are also shown (orange dashed lines). Plot generated with {\texttt{matplotlib}}~\cite{Hunter:2007}.}
\label{fig1:reach}
\end{figure}

Figure~\ref{fig1:reach} shows the $95\%$ CL reach at the 13 TeV LHC in the mass plane $(m_{W_{R}}, m_{N})$, based on signal sensitivity to 3 events in 300 and 3000 fb$^{-1}$. The green region corresponds to the limit extracted from our past work in Ref.~\cite{Cottin:2018kmq}\footnote{The on-shell production cross-sections used in Fig. 4 of~\cite{Cottin:2018kmq} were underestimated by a factor of 3. This was corrected and a corrected version should appear soon. We thank Goran Popara for hinting this may come from charge combinations.}. When considering off-shell $W_{R}$ production, the reach goes up to $m_{W_{R}}=6250$ GeV for $m_{N}=21$ GeV, and $m_{N}=31$ GeV for $m_{W_{R}}=3250$ GeV at 300/fb.  With 3000/fb, we are able to reach higher than $m_{N} = 45$ GeV for $m_{WR} = 4500$ GeV, and $m_{W_{R}}=11500$ GeV for $m_{N}=37$ GeV.

We also overlay in Figure~\ref{fig1:reach} (with dashed orange lines) the estimated reach with standard leptons + missing transverse momenta searches. These were estimated for the {\it{invisible}} region in~\cite{Nemevsek:2018bbt}. Note that the projection with 3000/fb from Ref.~\cite{Nemevsek:2018bbt} was performed for a center of mass energy of 14 TeV, but we still show this for comparison. We see that at high luminosity, this displaced search strategy turns out to be in fact more competitive than searches for invisible decays.

\section{Conclusions}
\label{sec:Summary}

We have revisited LHC prospects with displaced vertices in the search
for light, long-lived sterile neutrinos. We focused on the
{\it{displaced}} region of the minimal left-right symmetric model,
where the long-lived neutrino has a mass below the electroweak
scale. By considering that off-shell production of a $W_{R}$ dominates
at $\mathcal{O}$(GeV) neutrino masses (as shown in
Ref.~\cite{Nemevsek:2018bbt}), we show that the high luminosity LHC is
able to probe masses up to $\sim 45$ GeV and 11.5 TeV for sterile neutrinos and $W_{R}$
bosons, respectively. This newly calculated reach turns out to be
more competitive than the one projected with standard lepton + missing
energy searches. This becomes of great importance when assessing the
complementarity of different strategies, as backgrounds for more
standard searches will continue to grow at higher center of mass
energies, while for displaced vertex searches they will remain
low. Further improvements and optimizations of current displaced
searches at the LHC look promising for searches for light sterile
neutrinos in the left-right model.

\acknowledgments{We thank Goran Popara, Miha Nemevšek and Fabrizio Nesti for clarifying comments. 
  We thank Goran Popara for providing
  a working version of the KS event generator and help and advice with
  its usage. We thank Richard Ruiz for early correspondence
  related to this work. G.C. would also like to thank the Physics and
  Astronomy Department at University of La Serena for hospitality
  offered while working on this project. G.C. is supported by the
  Ministry of Science and Technology of Taiwan under grant
  No. MOST-107-2811-M-002-3120. J.C.H. is supported by Chile grant
  Fondecyt No. 1161463. M. H. was funded by Spanish MICINN grant
  FPA2017-85216-P and SEV-2014-0398 (from the Ministerio de Economía,
  Industria y Competitividad), as well as PROMETEO/2018/165 (from the
  Generalitat Valenciana).}

\FloatBarrier

\bibliographystyle{apsrev4-1}
\bibliography{main}% Produces the bibliography via BibTeX.

\end{document}